\documentclass[sn-nature]{sn-jnl}% Style for submissions to Nature Portfolio journals

\usepackage{graphicx}%
\usepackage{multirow}%
\usepackage{amsmath,amssymb,amsfonts}%
\usepackage{amsthm}%
\usepackage{mathrsfs}%
\usepackage{anyfontsize}          %% Manually added package

\usepackage[dvipsnames]{xcolor}
\usepackage{textcomp}%
\usepackage{manyfoot}%
\usepackage{booktabs}%
\usepackage{subcaption}
\usepackage{tikz}
\usetikzlibrary{shapes.geometric}
\usepackage{xparse}
\usepackage{forloop}
\usepackage{pifont}
\usepackage{booktabs} % For better-looking tables
\usepackage{siunitx}  % For units like Pflops and MW

\newcounter{x}
\newcommand{\stars}[1]{%
    \forloop{x}{0}{\value{x} < #1}{%
        $\textcolor{blue}\bullet$
    }%
    \forloop{x}{#1}{\value{x} < 5}{%
        $\textcolor{Gray}\circ$
    }%
}

\begin{document}

\title[Article Title]{Shifting sands of hardware and software in exascale quantum mechanical simulations}

\author*[1]{\fnm{Ravindra} \sur{Shinde}}\email{r.l.shinde@utwente.nl}

\author[1]{\fnm{Claudia} \sur{Filippi}}\email{c.filippi@utwente.nl}

\author[2]{\fnm{Anthony} \sur{Scemama}}\email{scemama@irsamc.ups-tlse.fr}

\author*[3]{\fnm{William} \sur{Jalby}}\email{william.jalby@uvsq.fr}

\affil[1]{\orgdiv{MESA+ Institute for Nanotechnology}, \orgname{University of Twente}, \orgaddress{\street{P.O. Box 217}, \city{Enschede}, \postcode{7500 AE}, \state{Overijssel}, \country{The Netherlands}}}

\affil[2]{\orgdiv{Laboratoire de Chimie et Physique Quantiques (LCPQ)}, \orgname{Universit\'e de Toulouse (UPS) and CNRS}, \orgaddress{\street{118, route de Narbonne}, \city{Toulouse}, \postcode{31062}, \country{France}}}

\affil[3]{\orgdiv{Universit\'e Paris-Saclay}, \orgname{UVSQ}, \orgaddress{\street{9 Boulevard d’Alembert}, \city{Guyancourt}, \postcode{78280}, \country{France}}}

\abstract{The era of exascale computing presents both exciting opportunities and unique challenges for quantum mechanical simulations. Although the transition from petaflops to exascale computing has been marked by a steady increase in computational power, it is accompanied by a shift towards heterogeneous architectures, with graphical processing units (GPUs) in particular gaining a dominant role. The exascale era therefore demands a fundamental shift in software development strategies. This review examines the changing landscape of hardware and software for exascale computing, highlighting the limitations of traditional algorithms and software implementations in light of the increasing use of heterogeneous architectures in high-end systems. We discuss the challenges of adapting quantum chemistry software to these new architectures, including the fragmentation of the software stack, the need for more efficient algorithms (including reduced precision versions) tailored for GPUs, and the importance of developing standardized libraries and programming models.}

\maketitle

\section*{Table-of-contents summary}

\section*{[H1] Introduction}

Quantum mechanical (QM) simulations are crucial for understanding and predicting the behaviour of molecules and materials, offering insights that drive advancements in diverse fields from quantum chemistry to materials science. These simulations, which rely on solving the electronic Schr\"{o}dinger equation, are computationally demanding, and their accuracy is directly tied to the complexity of the system being modeled and the chosen level of theory. Efficient execution of QM simulations has traditionally been enabled by advances in high-performance computing (HPC), with a relatively predictable trajectory of hardware and software development. A few dominant hardware vendors provided stable tools and clear paths for scientists to take advantage of increasing computational power, primarily through more central processing unit (CPU) cores and higher clock speeds. This steady progress culminated in the petascale era~\cite{bader_2007,geist_2009,vetter_2017,dongarra_2001}. 

However, the shift to exascale computing  has disrupted this trajectory, representing not merely a quantitative leap with a thousand-fold increase in computational power, but a qualitative shift in the landscape of HPC~\cite{pedretti_2020,kogge_2022,sinha_2022,patrizio_2022,loh_2023}. Instead of simply scaling existing architectures, the exascale era is defined by heterogeneous architectures.  These architectures prominently feature graphical processing units (GPUs) alongside traditional CPUs. This mirrors the successful utilization of GPU power by artificial intelligence (AI) applications, thanks to the alignment of AI algorithms with GPU strengths and the development of specialized libraries~\cite{lu_2018,huerta_2020,gepner_2021,liang_2023,dongarra_2011,fiore_2018,richard_2018,gordon_2020,luo_2020,mcinnes_2021,matsuoka_2023}. This shift presents both exciting prospects and significant challenges for physicists and chemists performing QM simulations~\cite{geist_2015,vetter_2017,evans_2021}.

A major issue is the increasing fragmentation of the once reliable software stack, making it difficult to efficiently utilize these new, diverse hardware options~\cite{dongarra_2011,fiore_2018,richard_2018,gordon_2020,luo_2020,mcinnes_2021,matsuoka_2023}. Existing QM codes and algorithms, often optimized for CPU-based systems, are not readily transferable to GPU-accelerated environments. Addressing this requires substantial, coordinated efforts, often exceeding individual capabilities and requiring large community initiatives such as the US Exascale Computing Project~\cite{DOE-ECP}, EU Centers of Excellence~\cite{CoE}, and National Competence Centers~\cite{EuroCC}. The development and adoption of modular standardized libraries for key computational kernels will be crucial to simplify code development and ensure long-term sustainability~\cite{qmckl, lehtola_2023}.

This review explores the uneven path toward exascale QM simulations, examining the implications of the changing hardware landscape, particularly the rise of heterogeneous architectures. We will discuss how fragmentation of the software stack impacts the development and deployment of QM software, and highlight the need for new algorithms and programming models that can effectively exploit the capabilities of GPUs, including the use of reduced precision calculations. Concrete examples from a wide range of QM methodologies will illustrate these issues.

Finally, we argue that a successful transition requires a collaborative and multifaceted approach. This includes designing new algorithms optimized for heterogeneous architectures, promoting energy-efficient computing practices, and co-designing codes with stronger collaboration between academia and industry. By addressing these challenges, the field can fully leverage exascale computing~\cite{kowalski_2021,pototschnig_2021,yokelson_2022,gavini_2023,kim_2023,galvezvallejo_2023,corzo_2023,schade_2023}, and unlock new frontiers in materials design, molecular discovery, and a deeper understanding of the quantum world.

\section*{[H1] Quantum mechanical simulations}

The computational cost of QM simulations is closely tied to the number of basis functions used to represent the wave function, which scales with the size and complexity of the system.
Additionally, the choice of method for solving the electronic Schr\"{o}dinger equation determines both the computational scaling and memory requirements.
However, despite algorithmic and HPC advancements, 
QM simulations on the whole remain computationally demanding, even with the advent of exascale computing.

Mean-field-like methods, including Hartree--Fock (HF), Kohn--Sham density functional theory (Kohn--Sham DFT)~\cite{yu_2016}, and density functional tight-binding (DFTB)~\cite{spiegelman_2020}, rely on self-consistent iterations to optimize the density matrix and minimize the total energy.
In each iteration, the Fock or Kohn--Sham matrix is constructed using one- and two-electron integrals, followed by matrix diagonalization to update molecular orbitals.
These steps, particularly the construction and diagonalization of the Fock or Kohn--Sham matrix, represent the main computational bottlenecks.
The matrix diagonalization scales as $O(N^3)$, and the computation of two-electron integrals scales as $O(N^4)$, making direct storage (for localized basis sets) prohibitive for large systems.
To address this computational cost, sparsity is exploited, and on-the-fly computation of integrals is employed for non-zero matrix elements.
When these strategies are used, favourable scaling of mean-field-like methods makes them well-suited for large systems and molecular dynamics simulations.

Post-Hartree--Fock methods, such as configuration interaction~\cite{davidsherrill_1999}, coupled cluster and multi-configurational approaches~\cite{lyakh_2012}, achieve higher accuracy but at substantially higher computational costs.
These methods typically involve a four-index transformation of integrals to the molecular orbital (MO) basis, which scales as $O(N^5)$ but can be reduced to $O(N^4)$ using low-rank approximations such as density fitting~\cite{jung_2005} or Cholesky decomposition~\cite{pedersen_2024}.
The wave function is expressed in a basis of Slater determinants or configuration state functions (CSFs), whose size grows from $O(N^4)$ to $O(N!)$.
Computational bottlenecks include finding the lowest roots of the Hamiltonian eigenvalue problem.
The Hamiltonian matrix is generally too large to fit in memory, so iterative methods are used, such as Davidson’s algorithm~\cite{davidson_1975}, which is bound by sparse matrix--vector multiplications, with matrix elements evaluated on-the-fly using integrals expressed in the MO basis.
Coupled cluster methods involve solving nonlinear equations~\cite{bartlett_2007}, where tensor contractions dominate the computational cost~\cite{corzo_2023}.
Perturbative methods, such as M{\o}ller--Plesset perturbation theory (MP2), are well-suited for massive parallelism due to their non-iterative nature.
Combined with the use of low-rank approximations, they can be used to study large systems with a high level of accuracy~\cite{barca_2021_2}.

Kohn--Sham DFT and time-dependent DFT~\cite{burke_2005,casida_2012} are widely used approaches for electronic structure calculations in the ground and excited state, though their accuracy depends on the choice of the approximate exchange--correlation functional and on the nature of the excitation (single or double excitation, Rydberg state, or charge transfer, for example). Green's function approaches such as the GW and Bethe--Salpeter equation (BSE) methods are more expensive but can often provide more accurate band gaps than Kohn--Sham DFT and more accurate excitation energies than time-dependent DFT. However, doubly excited states and strongly correlated systems, for example, still require the use of wave function methods.

Quantum chemistry codes like GAMESS~\cite{code_gamess}, NWChem~\cite{code_nwchemx}, and Dalton~\cite{code_dalton} typically employ localized basis functions centred at the nuclear positions (Gaussian-type orbitals).
These basis sets have the advantage of a compact representation. However, the computation of two-electron integrals often constitutes a substantial computational bottleneck.
Developing efficient algorithms for integral evaluation remains an active area of research~\cite{reine_2012}.
Although Gaussian basis functions are traditionally used for molecular simulations, they can also be applied to periodic systems, as implemented in PySCF~\cite{code_pyscf1,code_pyscf2}, CRYSTAL~\cite{code_crystal}, and CRYSCOR~\cite{code_cryscor}, including in post-HF methodologies. Beyond this, versatile simulations of both molecular and condensed-matter systems are enabled by hybrid approaches that combine Gaussian and plane-wave basis sets, as in CP2K~\cite{code_cp2k}, or by using numerical atom-centered orbitals, as in FHI-aims~\cite{code_fhiaims}.

For periodic systems, plane-wave-based codes such as VASP~\cite{code_vasp}, Quantum Espresso~\cite{code_qe}, ABINIT~\cite{code_abinit} and CASTEP~\cite{code_castep} are often used.
Plane waves naturally accommodate periodic boundary conditions, making them ideal for studying bulk materials, surfaces and interfaces.
Their computational bottleneck lies in the reliance on fast Fourier transforms (FFTs) to switch between real and reciprocal space. Communication overhead in large-scale parallel FFTs poses challenges for exascale architectures.
Real-space grid-based codes such as Octopus~\cite{code_octopus}, GPAW~\cite{code_gpaw}, SPARC~\cite{code_sparc}, RMG~\cite{code_rmgdft}, DFT-FE~\cite{code_dftfe1,code_dftfe2}, and PARSEC~\cite{code_parsec}, provide an alternative approach by solving equations on a discrete grid.
This is particularly advantageous for non-periodic and localized systems, allowing adaptive mesh refinement to enhance accuracy in critical regions while reducing computational costs.
Despite their flexibility, these codes face their own challenges, primarily associated with efficient sparse matrix operations. Applying the sparse Hamiltonian matrix to the wave function demands highly optimized sparse matrix--vector multiplications and, for a very large scalability, sparse matrix algebra is a limiting factor before communication.

In periodic systems, wave function methods necessitate the use of a supercell to accurately account for electron--electron interactions. This approach significantly increases the size of the basis set, leading to substantial computational demands. Consequently, high-accuracy wave function calculations are rarely applied to periodic systems. An exception is the CC4S code~\cite{code_cc4s}, which enables periodic coupled cluster calculations and represents an important advance in this area.
In condensed systems, for the computation of electronic excited states, the GW and BSE methods are preferred \cite{GW_review1, GW_review2}. Notable examples include BerkeleyGW~\cite{code_berkeleygw}, Exciting~\cite{code_exciting}, and Yambo~\cite{code_yambo}. These specialized codes are typically used in conjunction with periodic DFT codes, using information from a preliminary ground-state calculation.

Real-space quantum Monte Carlo (QMC) methods, such as diffusion Monte Carlo, employ stochastic sampling to solve the many-body Schr\"{o}dinger equation~\cite{QMC_foulkes,QMC_lester1,QMC_lester2,QMC_mitas}.
These methods are inherently well-suited for massively parallel simulations and can achieve accuracies comparable to coupled cluster approaches.
QMC is applicable to both periodic systems and molecular systems, using either plane-wave or localized basis sets.
Its computational complexity and memory requirements are generally lower than those of other highly accurate wave-function-based methods, making QMC an attractive choice for larger systems.
However, QMC methods have a high computational prefactor arising from the need for extensive sampling and the evaluation of complex wave functions. Key computational bottlenecks include the evaluation of the wave function at electron positions and the optimization of the wave function. Unlike deterministic methods, these challenges cannot be reduced to a few large linear algebra operations. Instead, QMC calculations involve a sequence of diverse computational kernels that handle relatively small data sizes but require repeated execution on a massive scale.
Because of their specialized computational requirements, QMC methods are not typically implemented in mainstream quantum chemistry codes. Instead, they are supported by dedicated software packages such as QMCPACK~\cite{code_qmcpack1,code_qmcpack2}, TurboRVB~\cite{code_turborvb}, CHAMP~\cite{code_champ}, QMC=Chem~\cite{code_qmcchem}, QWalk~\cite{code_qwalk} and CASINO~\cite{code_casino}, which are interfaced with other codes to generate the trial wave functions needed for QMC calculations.
%
%---
Interestingly, the emergence of neural network-based wave functions has further expanded the scope of QMC methods. By making use of neural networks to parameterize the wave functions, these approaches can achieve highly accurate solutions to the electronic Schr{\"o}dinger equation~\cite{carleo_2017,ferminet,paulinet,neuralnet,deepwf}. This innovation highlights the synergy between machine learning and quantum mechanical simulations.

Tackling the inherent trade-offs between accuracy and efficiency has driven researchers to increasingly explore artificial intelligence (AI) and machine learning (ML) as complementary tools. 
For instance, machine learning potentials trained on \textit{ab initio} data can predict potential energy surfaces and forces with quantum-level accuracy at a significantly reduced computational cost~\cite{Gordon_Bell_Roberto_Car, car_NN,  kozinsky_nrp, kozinsky_ML_XC, borinsky_ML_FF, csanyi_ML_FF, csanyi_ML_GAP, ceriotti_ML_FF, ceriotti_atomistic_ML, Chandrasekaran2019}.
These advances have enabled efficient materials discovery and design through high-throughput calculations, supported by public datasets~\cite{materialsproject} and advanced optimization techniques~\cite{MLHIFI,batatia_2023}.

\section*{[H1] The transition to exascale}
The theoretical maximum speed of a computer is given by the total number of cores multiplied by their nominal speed. However, the actual speeds achieved can be far lower and need to be measured using benchmarks built on real applications with realistic data sets.
The progress in HPC is marked by milestones tracked by the Top500 list~\cite{top500}, which monitors peak computational power through a benchmark known as LINPACK~\cite{linpack}. It measures the performance of systems in solving a dense linear system, a task optimized for high computational throughput with minimal data movement and communication.
Remarkably, LINPACK achieves around 70\% of the theoretical maximum speed. 
We have witnessed steady progress in computational power, culminating in crossing the exaflop barrier some time in 2022.

However, LINPACK’s numbers, though impressive, are not fully representative of real-world performance. A more challenging and realistic benchmark is the High-Performance Conjugate-Gradient (HPCG) benchmark, which evaluates the ability of a computer to solve sparse linear systems and is therefore representative of a different type of numerical applications, for instance, based on finite elements and finite difference schemes~\cite{hpcg}. The performance in this test is starkly different, typically approximately 20 times lower than LINPACK, using less than 5\% of the nominal speed~\cite{top500-hpcg}. This contrast underscores the difference between idealized benchmarks and practical workloads.

Although LINPACK and HPCG represent reasonable upper and lower bounds of performance, neither reflects the complexity of actual applications. The Gordon Bell Prize, awarded for outstanding achievements in HPC applications, provides a more relevant measure~\cite{gordon-bell}. These applications, often solving real scientific problems, have demonstrated performance nearing the exaflop threshold~\cite{liu_2021, gordon_finalist, gordon_finalist2, gordon_winner}. Achieving this requires monumental efforts: entire codebases are rewritten by interdisciplinary teams of physicists, chemists, mathematicians and computer scientists. New algorithms are developed, and implementations are tailored to exploit the specific features of the target architectures.

In contrast, many legacy codes, which comprise the bulk of real-world applications, lag far behind. Many such codes were not designed with exascale capabilities in mind and thus fall short of the exaflop performance. The challenge ahead lies in bridging this gap, ensuring that applications needing exascale capabilities can harness the full potential of exascale systems.

\section*{[H1] The changing hardware landscape}

The transition from petaflops to exascale computing marked a significant departure from the consistent and predictable trajectory of high-performance computing. Whereas the climb to petaflops was characterized by steady increases in processor core count and clock speeds, exascale computing brings a fundamental shift in hardware architectures and software development strategies.
One of the most significant changes is the increasing adoption of heterogeneous architectures. Unlike the homogeneous central processing unit (CPU)-based systems that dominated petaflops, exascale systems heavily rely on accelerators like GPUs, which offer massive parallelism for specific tasks but come with unique challenges. This heterogeneous landscape presents a major hurdle for software developers as traditional algorithms and software stacks designed for CPUs need significant modifications to utilize the capabilities of GPUs efficiently.

Before exploring the hardware evolution, it is crucial to identify the primary application areas driving advancements in hardware, as they fundamentally determine the level of investment and the trajectory of hardware development. Scientific HPC, including QM applications, remains a niche market and does not significantly influence major hardware innovations. Instead, scientific computing and exaflop architectures often repurpose software and hardware technology developed for other, larger markets.
Two key segments, that are increasingly shaping the practices of the scientific community, are cloud and accelerator computing, driven primarily by their adoption in conventional companies and the field of AI, respectively. The general-purpose laptop segment also indirectly impacts high-end processor computing. For instance, Apple, a major customer of the Taiwan Semiconductor Manufacturing Company (TSMC), exerts significant pressure on high-end chip production for both cloud computing and AI applications due to their high volume demand, as these components share the same fabrication lines~\cite{tsmc-apple}.

In analyzing hardware evolution, we focus on three major components: CPUs, accelerators, and supercomputing systems.

On the CPU front, changes have been incremental over the past decade. Major CPU manufacturers (AMD, Intel, ARM) have converged towards a similar generic architecture: multicore processors based on out-of-order and superscalar technology. An important development has been the increase in the number of cores per processor, now exceeding 100. Although other technologies have emerged, none have achieved widespread adoption. Wider vectors (up to 512 bits) have been embraced by x86 architectures (initially by Intel, followed by AMD), whereas ARM continues to use shorter vectors (128 or 256 bits) without major performance penalties. Notably, memory technology has evolved more significantly, with very large level 3 caches using 3D stacking and High Bandwidth Memory (HBM) offering an alternative to the standard double data rate synchronous dynamic random access memory (DDR-SDRAM). The performance gains from these new memory technologies depend heavily on data access characteristics of the application.

\begin{table}
\renewcommand{\arraystretch}{1.5}
\caption{\label{tab:comparison} \textbf{Comparison of the peak performance} with different precision modes, the amount of memory, the memory bandwidth, and the thermal design power of an Nvidia H100 GPU and an AMD EPYC (Bergamo) CPU. FP, floating point; HBM, high bandwidth memory; DDR, Double Data Rate; TDP, thermal design power.  \textsuperscript{a}Data from Ref.~\cite{data_nvidia}. \textsuperscript{b}Data from Ref.~\cite{data_amd}.}
\begin{tabular}{lcc}
\hline
\textbf{Processing unit}    &  \textbf{Nvidia H100 SXM}\textsuperscript{a}  & \textbf{AMD EPYC\texttrademark 
 \phantom{0}9754}\textsuperscript{b} \\
Cores                       &  16896    & 128 \\
Clock speed                 &  1.6 GHz  & 2.25 GHz \\
\hline
Release date       &  September 2022    &  June 2023 \\
\hline
FP64               & \phantom{00}34~TFlop/s   & \phantom{0}6.9~TFlop/s\\
FP64 (tensor core) & \phantom{00}67~TFlop/s   & -- \\
FP32               & \phantom{00}67~TFlop/s   & 13.8~TFlop/s \\
FP32 (tensor core) & \phantom{0}989~TFlop/s  & -- \\
FP16 (tensor core) & 1979~TFlop/s & -- \\
FP8  (tensor core) & 3958~TFlop/s & -- \\
\hline
Memory             & 80~GB             &  max 6000~GB \\
\hline
Memory Bandwidth   & 3350~GB/s (HBM3)        & 460.8~GB/s (DDR5) \\
\hline
Interconnection Nvlink &  900~GB/s & \\
Interconnection PCIe &  \multicolumn{2}{c}{128~GB/s} \\
Quantum X800 Infiniband network   & 100~GB/s & 100~GB/s \\
\hline
TDP                & 700~W             &  360~W \\
\hline
\end{tabular}
\end{table}

On the accelerator front, the primary advance has been the massive increase in core count, now reaching thousands --- nearly two orders of magnitude more than CPUs. The number of cores and peak performance depend on the floating-point (FP) format; typically, double-precision (DP) cores are half as numerous as single-precision (SP) cores. The shift toward narrower FP formats has been driven by AI, which tolerates reduced precision (16-bit, 8-bit, and even 4-bit formats are in use). Modern GPUs achieve remarkable performance using smaller FP formats and Tensor Cores, which are optimized for dense matrix multiplications.
To illustrate these differences, Table~\ref{tab:comparison}  compares the computational performance of a high-performance CPU and a GPU released within the same time frame.
Peak performance reflects execution speed only when computation is the bottleneck, not memory transfers or inter-node communications, which are too slow to keep up with the execution units.
With standard DP operations (FP64), GPU peak performance is about five times that of a high-end CPU. When Tensor Cores are used, this increases to a factor of ten. For SP (FP32) operations, both CPU and GPU performance doubles, but with Tensor Cores, the GPU reaches 989 TFlop/s, over 70 times that of the CPU. Reducing precision to 8 bits (FP8) adds another factor of four.
Whereas smaller FP formats can sometimes be used for standard FP32 and FP64 tasks, Tensor Cores are mainly of benefit to computations reliant on dense matrix multiplication.
Achieving GPU peak performance is harder than on CPUs, as it requires high levels of parallelism. Consequently, matrices on the GPU must be much larger~\cite{sorokin_2022},  or numerous, requiring batching algorithms. Furthermore, GPUs have less memory than CPUs~\cite{amd_memory} and large workloads necessitate memory transfers from the host to the accelerator, which must be managed in parallel with computations to prevent communication bottlenecks.
We further elaborate below on the implications of these CPU and GPU performance differences for quantum simulations.

Finally, on the supercomputing systems front, the changes have been rather radical. Using the Top500 list as a reference, most of the major changes in the supercomputing domain can be readily monitored. First, the amount of parallelism has drastically increased: the top runner in Top500 had 200,000 cores in June 2010, and now it has 11 million cores (1 million CPU and 10 million GPU cores; a 55$\times$ increase). The overall system organization has also radically changed: in 2010, GPUs were used in fewer than 40 systems among the 500, now over half of the systems are equipped with GPUs.
Today's top systems exhibit a heterogeneous architecture, integrating both CPUs and GPUs, though the bulk of computational capability predominantly stems from the GPUs. Moreover, the total power consumption of these systems has increased by an order of magnitude, rising from 3~MW to 30~MW, albeit with a notably irregular progression~\cite{top500}. This increase underscores the substantial enhancements in system design, especially in terms of energy efficiency. This situation might change with the upcoming development of efficient processors and accelerator technology based on open-standard RISC-V instruction set architecture~\cite{riscv}.

\section*{[H1] Increased fragmentation in the software stack}

The transition from petascale to exascale computing has brought unprecedented computational power but has also introduced significant challenges in adapting software to efficiently utilize the new architectures.
One of the most prominent challenges is the increasing fragmentation of the software stack, particularly with GPU programming. Nvidia had an early advantage for being a pioneer in this area. This created a relatively unified ecosystem where developers could rely on a single set of tools and libraries to harness the power of GPUs. However, as GPU computing gained wider adoption, other vendors such as AMD and Intel entered the market, each with their own programming models (HIP and SYCL, respectively). This proliferation of models has led to a fragmented landscape, making it increasingly difficult for developers to write portable and efficient code that can run seamlessly across different GPU platforms or heterogeneous systems.

To address portability, OpenCL was introduced. It provides an open standard for
CPUs, GPUs and field programmable gate arrays (FPGAs) across vendors. However, it has not gained wide adoption
in HPC due to performance and usability issues. Although OpenCL offers portability,
its abstraction often results in suboptimal performance compared to specialized
frameworks like CUDA \cite{openCL_portability}. Its lower-level programming model is also cumbersome,
requiring substantial manual effort for memory management and kernel
optimization. Fragmented implementations across vendors further undermine
OpenCL's portability, as performance and features can vary, forcing developers to
fine-tune code for each platform.

The lack of an efficient standardized programming model for GPUs has resulted in a number of issues for developers. Firstly, it creates vendor lock-in, where codes written for one platform cannot easily be migrated to another without substantial code refactoring.
For example, adopting Nvidia programming models involves considerable trade-offs between performance and portability across different hardware environments~\cite{GTC22conf}. Another issue stemming from the fragmentation is the lack of compiler support for all programming models on all platforms. For instance, (Table \ref{tab:vendor_support}), whereas CUDA is well supported on Nvidia GPUs through the CUDA Toolkit, its support on AMD and Intel hardware is limited and relies on indirect means like translation layers or third-party libraries and utilities~\cite{syclomatic,dpcpp,hipify,zluda}. This forces developers to rely on vendor-specific compilers, which often lack the universality and feature-richness of their CPU counterparts.

    \begin{table}[h!]
        \centering
        \renewcommand{\arraystretch}{1.5}
        \begin{tabular}{l|lll|lll}
        \hline
        \multirow{2}{*}{\textbf{Tool}} & \multicolumn{3}{c|}{\textbf{C++}}              & \multicolumn{3}{c}{\textbf{Fortran}}          \\
        & \textbf{Intel} & \textbf{AMD} & \textbf{Nvidia} & \textbf{Intel} & \textbf{AMD} & \textbf{Nvidia} \\ \hline
        OpenACC  & \stars{1} & \stars{3} & \stars{5} & \stars{1} & \stars{2} & \stars{5} \\
        OpenMP   & \stars{5} & \stars{5} & \stars{4} & \stars{5} & \stars{5} & \stars{5} \\
        Standard & \stars{5} & \stars{1} & \stars{5} & \stars{5} & \stars{0} & \stars{5} \\
        CUDA     & \stars{4} & \stars{4} & \stars{5} & \stars{0} & \stars{1} & \stars{5} \\
        HIP      & \stars{3} & \stars{5} & \stars{4} & \stars{0} & \stars{1} & \stars{1} \\
        SYCL     & \stars{5} & \stars{3} & \stars{3} & \stars{0} & \stars{0} & \stars{0} \\
        Alpaka   & \stars{3} & \stars{3} & \stars{3} & \stars{0} & \stars{0} & \stars{0} \\
        Kokkos   & \stars{3} & \stars{3} & \stars{3} & \stars{1} & \stars{1} & \stars{1} \\ \hline
        \end{tabular}
        \caption{\textbf{Vendor support compatibility on GPUs.} 
\stars{5}: Full support with complete implementation, documentation, updates, and error handling;
\stars{4}: Indirect but comprehensive support via translation to a native model;
\stars{3}: Partial vendor support; most features work, but some may be unavailable;
\stars{2}: Comprehensive support exists but is community-driven, not vendor-provided;
\stars{1}: Limited support requiring significant user effort;
\stars{0}: No direct support; workarounds like custom headers or manual linking may be needed;
\textsuperscript{a}Denotes built-in parallelization by compilers. Table data from Ref.~\cite{Herten2023}.}
\label{tab:vendor_support}
\end{table}

The compatibility between various GPU programming models and vendors is complex, largely due to the growing number of choices involving GPU platforms, programming models, and languages. As recently analyzed by Herten~\cite{Herten2023}, while OpenMP is natively supported across the three major platforms—AMD, Intel, and Nvidia—and works with both C\texttt{++} and Fortran, other popular models like OpenACC have limited support on Intel GPUs. This underscores the importance of thoroughly assessing a programming model's level of support on a chosen platform before beginning code development. Additionally, standardized benchmarks and evaluation tools are needed to effectively compare the performance and capabilities of different programming models.

The lack of standardization in GPU programming has also led to a variety of community-driven, higher-level models that aim to abstract away vendor-specific details and provide a more portable programming experience. Examples of such models include Kokkos~\cite{kokkos}, RAJA~\cite{raja},  Alpaka~\cite{alpaka}, and, to a lesser extent, hipSYCL~\cite{hipsycl} which is limited to AMD and Nvidia. These abstraction models often use vendor-native infrastructure in the background, enabling developers to write code that can be deployed on multiple platforms without substantial code changes. However, the support and standardization of these higher-level models can vary greatly, and relying on community-driven efforts for critical software infrastructures can introduce uncertainties and complexities in the long run.
Comparing the performance of various GPU programming models on the LUMI supercomputer~\cite{lumi_gpu_benchmark} reveals the potential of community-driven models like Kokkos in achieving portable performance but also highlights the need for further development and community support to ensure their long-term viability.

Another important aspect of the software stack is the role of low-level libraries for key computational kernels, such as linear algebra, FFTs, and communication routines. These libraries play a critical role in achieving high performance and efficiency on exascale systems. However, the use of different libraries across different platforms can also contribute to software fragmentation. For example, whereas Nvidia provides the highly optimized cuBLAS library for linear algebra operations on their GPUs, AMD offers the RocBLAS library. These libraries, despite providing similar functionality, often have different APIs and performance characteristics, further hindering code portability.

The exascale era has reinforced the need for a more unified and standardized software stack for GPU programming. Fragmentation creates barriers to code portability, hinders performance optimization, and increases the development effort required for scientific applications. As the community moves forward, it must address these challenges. Initiatives like SYCL, OpenMP, and community-driven efforts like Kokkos and RAJA look promising toward achieving a more unified programming experience. However, vendors, developers, and researchers must collaborate closely to establish common standards and standardize a more cohesive and interoperable software ecosystem for GPU computing.

\section*{[H1] QM and HPC}

Each QM method relies on distinct computational techniques (such as FFTs, matrix--vector operations or diagonalization of sparse or dense matrices). The techniques determine the scaling behaviour, whereas the specific hardware implementation governs the prefactor associated with this scaling. For instance, matrix multiplication kernels are central to coupled cluster methods that utilize large tensors, enabling efficient use of tensor cores and resulting in a relatively small prefactor despite the high scaling complexity. Conversely, although QMC methods are inherently massively parallel, they often operate on relatively small matrices, leading to a large prefactor; therefore, scaling these approaches on accelerators requires substantial effort~\cite{code_qmcpack2}.

Advances in computational chemistry have increasingly used GPUs to accelerate a range of quantum chemical methods.
For Hartree–Fock and DFT calculations, hybrid CPU/GPU implementations have demonstrated notable speedups. 
Early efforts in this area, such as the development of DFT calculations on multicore and multithreaded hybrid CPU/GPU approaches to Hartree–Fock \cite{genovese_2009, asadchev_2012_2}, were pioneering for their time but relied on significantly less powerful GPU hardware than current-generation GPUs.
More recent implementations taking advantage of modern GPU architectures have indeed demonstrated even greater performance gains \cite{manathunga_2020, barca_2020_2}. GPU-accelerated self-consistent field methods have been optimized for large-scale molecular simulations~\cite{stone_2007, seritan_2020, seritan_2021}.
Coupled-cluster methods, particularly in their density-fitted and non-iterative forms, have also been adapted for GPU architectures to address both single- and multi-reference systems~\cite{ma_2011, deprince_2011, deprince_2014, bhaskaran-nair_2013, kowalski_2021, shen_2019}. 
Additionally, divide-and-conquer approaches have been employed to scale electronic structure calculations for large systems \cite{wang_2014, hoyvik_2012, kjaergaard_2017}. Real-space grid-based methods offer interesting advantages as they are FFT-free (requiring fewer message passing interface (MPI) communications) and have local operations, and therefore can be massively parallelized. 

Beyond traditional methods, GPU-accelerated phaseless auxiliary-field quantum Monte Carlo~\cite{shee_2018} and multi-GPU configuration interaction~\cite{fales_2020} techniques have expanded the computational toolkit for tackling strongly correlated systems.
Finally, specific applications, such as accelerating plane-wave-based calculations in VASP~\cite{hacene_2012} and implementing GPU-based GW calculations~\cite{delben_2020}, further highlight the versatility of GPU acceleration in computational chemistry. 

The efficient evaluation of electron repulsion integrals on GPUs has become a central focus in molecular quantum chemistry. Developments in this area offer a great deal of potential to accelerate electronic structure calculations.
Early foundational efforts~\cite{yasuda_2008, ufimtsev_2008} introduced GPU-based approaches to integral evaluation, which have since been refined through a variety of algorithms and optimization strategies, including the use of mixed-precision techniques~\cite{asadchev_2010, asadchev_2012, luehr_2011}.
Subsequent developments have addressed challenges such as higher angular momentum integrals~\cite{miao_2013, miao_2015, asadchev_2023}, algorithmic innovations like the BRUSH algorithm~\cite{rak_2015}, and automated code generation for GPU-based integral evaluation~\cite{song_2016}.
Hybrid CPU/GPU approaches~\cite{kussmann_2017} and advancements in GPU compiler technologies~\cite{tornai_2019} have further expanded the computational possibilities.
The use of multinode, multi-GPU systems for large-scale calculations has been recently demonstrated~\cite{johnson_2022}, as well as high-performance integral evaluation implementations~\cite{galvezvallejo_2023_2}.

The adoption of mixed single and double precision has been explored at various stages in computational methods.
Initially, this approach was investigated at the level of linear system solvers~\cite{buttari_2008}.
Subsequently, it gained traction in quantum chemistry methods, including MP2~\cite{vysotskiy_2011,olivares-amaya_2010,mixed_precision_iterative_diag}, QMC simulations~\cite{scemama_2013}, coupled-cluster methods~\cite{pokhilko_2018} and quantum molecular dynamics~\cite{mixed_precision_linear_scaling}.
The integration of single precision was first introduced in early GPU-accelerated quantum chemistry codes, notably through optimized matrix multiplication libraries~\cite{olivares-amaya_2010} and then in the acceleration of two-electron integral computations~\cite{asadchev_2010,luehr_2011,asadchev_2012,asadchev_2012_2}.
These implementations provided substantial speedups by leveraging the performance advantages of single-precision arithmetic on GPUs.
The use of mixed precision for the solution of the non-linear Kohn--Sham eigenvalue problem in the DFT-FE code resulted in the execution time being reduced by approximately a factor of two compared to double precision for a massively parallel calculation~\cite{code_dftfe1}.
Furthermore, the use of single precision in initial iterations of eigensolvers speeds up calculations on a massive scale without compromising final accuracy~\cite{mixed_precision_elpa}. 

More recently, research has extended to the use of half-precision (16-bit floating point) as a means of further optimizing computational performance while maintaining an acceptable level of accuracy.
Its potential benefits and challenges have been explored in real applications~\cite{abdelfattah_2019, poulos_2022}, including in the context of quantum chemistry~\cite{dawson_2024}, where accurate approximations for matrix products using fast and error-free splitting of floating-point numbers~\cite{Ozaki_Rump_scheme} were explored.
Mixed precision strategies in the WEST code achieved significant speedups for large-scale, full-frequency GW calculations \cite{code_west}.  Furthermore, calculations of exchange interactions in periodic systems using single-precision arithmetic, particularly within hybrid DFT, have demonstrated substantial reductions in computation time and memory requirements with negligible loss of accuracy in band energies, forces, and x-ray absorption spectra \cite{single_precision_exx}.

Efforts to adapt to the heterogeneous architectures of HPC systems highlight the critical importance of balancing performance with portability. Some codes, such as ABINIT and Quantum Espresso, have adopted hardware-agnostic offload programming models (OpenMP or OpenACC), whereas others, such as CP2K, have opted for non-portable, kernel-based models (CUDA and HIP). BerkeleyGW provides both CUDA and OpenACC alternatives, with the latter achieving performance within ten percent of the former~\cite{delben_2020}. Additionally, different communities are actively developing specialized performance libraries to execute computationally intensive tasks on accelerators. Examples include the Distributed Block Compressed Sparse Row Matrix (DBCSR) library~\cite{dbcsr} for Nvidia and AMD GPUs and the LibintX~\cite{libint, libintx} library, which provides efficient implementations of Gaussian integral evaluation, central to integral-direct implementations. 
The QM community library projects such as ELSI~\cite{elsi}, and especially ELPA, have proved very efficient and have been adopted by many codes\cite{elpa}.
The QMCkl library, developed within the TREX CoE, offers high-performance computation kernels for QMC simulations using OpenMP and OpenACC~\cite{qmckl}. The wave-function community~\cite{code_cc4s,pototschnig_2021,code_nwchemx,code_peps_torch,code_itensor,menczer_2024} has been extensively developing tensor libraries that automatically parallelize and accelerate operations on multidimensional arrays, taking advantage of modern HPC architectures~\cite{tiledarray, cyclops, tblis}. 

Although many studies have reported improved timings for QM calculations on heterogeneous architectures, the energy-to-solution costs remain a relatively unexplored aspect. Also, in this regard a direct comparison of timings with CPU-only systems can be misleading because these architectures differ significantly in power consumption, memory bandwidth, and hardware complexity, all of which should be accounted for. For example, an energy-efficient distributed, block-cyclic setup of the Hamiltonian and overlap matrices for linearized augmented plane wave calculations has been reported \cite{exciting_cpu_gpu}. Another large-scale benchmark study on the Summit supercomputer --- the nodes of which house two IBM POWER9 CPU sockets and six NVIDIA V100 GPUs connected via high-bandwidth NVLink --- further underscores the architectural disparities \cite{Gordon_Bell_Roberto_Car}. When normalized under equivalent power usage, the GPU-accelerated system provided up to sevenfold speedups compared to CPU-only (that is, using only the CPUs from these heterogeneous nodes) system. 

\begin{table}[t]
\centering
\caption{\textbf{Comparison of supercomputer performance and energy consumption across different benchmarks}. HPL values represent peak performance while HPCG illustrates more realistic workload performance. 
}
\label{tab:hpcg}
\begin{tabular}{@{}lccS[table-format=2.1]S[table-format=4.0]S[table-format=2.1]@{}}
\toprule
\textbf{Supercomputer} & \textbf{Year} & \textbf{Configuration} & \textbf{Power} & \textbf{HPL} & \textbf{HPCG} \\
 & & & \text{(MW)} & \text{(Pflops)} &  \text{(Pflops)} \\
\midrule
Fugaku   & 2020 & CPU-only & 29.9  & 442  & 16.0 \\
Frontier & 2021 & CPU/GPU  & 24.6  & 1353 & 14.0 \\
Aurora   & 2024 & CPU/GPU  & 38.7  & 1012 & 5.6  \\
\bottomrule
\end{tabular}
\end{table}

However, it is crucial to interpret such results with caution. This type of comparison, while isolating the contribution of the GPU, does not represent a realistic operational scenario. Furthermore, measuring the speedup of a GPU should not be done by comparison with the CPUs hosting those GPUs, but by a direct comparison to a separate, dedicated CPU-only system, which typically employs better-performing, higher-end CPUs.
Therefore, these results illustrate that although heterogeneous systems can deliver remarkable performance gains, their inherently different hardware profiles preclude an entirely fair one-to-one comparison with CPU-only configurations without simultaneously accounting for power requirements, memory constraints and scaling behaviour (see Table~\ref{tab:hpcg}).

\section*{[H1] Outlook}

The future of QM simulations at the exascale level hinges on effectively navigating the shifting landscape of HPC hardware and software. A critical aspect of this adaptation is the development of algorithms that go beyond simply exploiting the brute force of massive parallelism provided by GPUs. Although GPUs offer substantial acceleration,
the impressive performance gain is limited to algorithms that use matrix--matrix multiplication with reduced precision.
Designing alternative algorithms for QM simulations that make use of this particular strength of GPUs is a must.
Furthermore, given the complexity of adapting legacy codes to hybrid architectures, the community must unite in embracing and standardizing modular libraries for quantum chemical calculations, providing reusable, optimized and versatile implementations of complex algorithms~\cite{lehtola_2023}.
Some specialized libraries have already succeded significant success in enabling efficient implementations on diverse hardware platforms, and their continued development and community-wide adoption are crucial for simplifying code development and ensuring long-term code sustainability.
Finally, AI has already proven its broad potential, generating machine-learning force fields with \textit{ab initio} accuracy, faithfully describing quantum states and speeding up various tasks within quantum simulations~\cite{kulik_2022}. As a result, the growing integration of AI with quantum simulations offers a promising path for future breakthroughs.

However, a few notes of caution are necessary regarding the widespread adoption of GPUs.
CPUs remain essential for workloads that require flexibility, general-purpose computation, or that are memory-bound. For example, CPUs equipped with HBM are competitive with GPUs in some cases, particularly in memory-limited tasks. Although in the near future, we can expect CPUs to integrate AI acceleration features, they will likely remain complementary to GPUs.
In addition, although their computational power is undeniable, the energy consumption of GPUs is a growing concern. Whereas a laptop requires less than 100W to function, the peak energy consumption of a modern supercomputer like El Capitan is around 29MW. This stark contrast, while not intended as a direct performance comparison, highlights the immense difference in scale and the critical need for energy-efficient exascale computing. Even though the total energy consumption of exascale systems will inevitably be higher than previous generations due to their increased scale and capabilities, the community must prioritize energy efficiency as one of the central design principles for future exascale simulations. Time-to-solution should be accompanied by energy-to-solution when comparing simulations on different architectures. Re-evaluating the blind pursuit of raw performance at the cost of enormous energy consumption is of paramount importance. The focus should shift towards a more balanced approach in which algorithmic efficiency, software optimization and energy-aware hardware design work in synergy.

Importantly, although the exascale era offers exciting prospects for QM simulations, it also demands a cohesive effort from the scientific community to overcome the associated challenges. Collaborative initiatives, such as those driven by global research collaborations and industry partnerships, will be vital in shaping the future of computational chemistry and materials science. By embracing energy-efficient computing, stronger collaboration between academia and industry, and prioritizing the development of standardized software tools (such as languages, compilers, debuggers and libraries), we can ensure that exascale computing delivers on its promise as a cornerstone of scientific innovation, enabling us to take on grand challenges and push the boundaries of our understanding of the quantum world.

\backmatter

\bibliography{main}

\section*{Acknowledgments}
The authors acknowledge partial support from the European Centre of
Excellence in Exascale Computing TREX --- Targeting Real Chemical
Accuracy at the Exascale. This project has received funding in part from the
European Union's Horizon 2020 --- Research and Innovation Program ---
under grant agreement no.~952165.

\section*{Author contributions}
All authors contributed to all aspects of this work.

\section*{Competing interests}
The authors declare no competing interests.

\section*{Peer review information}

\end{document}